\documentclass{article}

\usepackage[preprint]{neurips_data_2023}

\usepackage[neverdecrease]{paralist}
\usepackage[utf8]{inputenc} 
\usepackage[T1]{fontenc}    
\usepackage{hyperref}       
\usepackage{url}            
\usepackage{booktabs}       
\usepackage{amsfonts}       
\usepackage{nicefrac}       
\usepackage{microtype}      
\usepackage{xcolor}         
\usepackage{graphicx}       
\usepackage{multirow}

\graphicspath{{figs/}} 
\usepackage{array}
\usepackage{tabularx}
\usepackage{float}
\usepackage{caption}
\usepackage{subcaption}  
\usepackage{amsmath}  
\usepackage{amsfonts}
\usepackage{amssymb}   
\usepackage{amsthm}
\usepackage{thmtools}
\usepackage{indentfirst} 
\usepackage{hyperref}
\usepackage{wrapfig}
\usepackage{makecell}

\newcommand\MyBoxOrg[2]{
\centering
  \fbox{\lower0.75cm
    \vbox to 1.7cm{\vfil
      \hbox to 1.7cm{\hfil\parbox{1.4cm}{#1\\#2}\hfil}
      \vfil}%
  }%
}

\newcommand\MyBox[2]{
\centering
  \fbox{\lower0.5cm
    \vbox to 0.8cm{\vfil
      \hbox to 0.8cm{\hfil\parbox{0.8cm}\centering{#1}\hfil}
      \vfil}%
  }%
}

\title{The Open Review-Based (ORB) dataset: Towards Automatic Assessment of Scientific Papers and Experiment Proposals in High-Energy Physics}

\author{%
  \textbf{Jaroslaw Szumega\textsuperscript{1,2,}\thanks{Corresponding author}
 \quad Lamine Bougueroua\textsuperscript{3} \quad Blerina Gkotse\textsuperscript{1,2}}\\
  \textbf{Pierre Jouvelot\textsuperscript{2}} \quad \textbf{Federico Ravotti\textsuperscript{1}} \\
  \textsuperscript{1}Experimental Physics Department, CERN, Geneva, Switzerland\\
  \textsuperscript{2}Mines Paris, Universit\'e PSL, Paris, France\\
  \textsuperscript{3}Efrei Research Lab, Universit\'e Paris-Panth\'eon-Assas, Paris, France
}
\begin{document}

\maketitle

\begin{abstract}
With the Open Science approach becoming important for research, the evolution towards open scientific-paper reviews is making an impact on the scientific community. However, there is a lack of publicly available resources for conducting research activities related to this subject, as only a limited number of journals and conferences currently allow access to their review process for interested parties.
In this paper, we introduce the new comprehensive Open Review-Based dataset (ORB); it includes a curated list of more than 36,000 scientific papers with their more than 89,000 reviews and final decisions. We gather this information from two sources: the \mbox{\textit{OpenReview.net}} and \mbox{\textit{SciPost.org}} websites. However, given the volatile nature of this domain, the software infrastructure that we introduce to supplement the ORB dataset is designed to accommodate additional resources in the future. 
The ORB deliverables include (1) Python code (interfaces and implementations) to translate document data and metadata into a structured and high-level representation, (2) an ETL process (Extract, Transform, Load) to facilitate the automatic updates from defined sources and (3) data files representing the structured data.
The paper presents our data architecture and an overview of the collected data along with relevant statistics. For illustration purposes, we also discuss preliminary Natural-Language-Processing-based experiments that aim to predict (1) papers' acceptance based on their textual embeddings, and (2) grading statistics inferred from embeddings as well.
We believe ORB provides a valuable resource for researchers interested in open science and review, with our implementation easing the use of this data for further analysis and experimentation. We plan to update ORB as the field matures as well as introduce new resources even more fitted to dedicated scientific domains such as High-Energy Physics.
\end{abstract}

\section{Introduction}
\label{sec:introduction}
The motivation for the introduction of the ORB dataset dedicated to the reviews of scientific publications stems from a quite specific research goal. Within the framework of the RADNEXT project\footnote{\url{https://radnext.web.cern.ch}} which sponsors international collaborations for the performance of High-Energy Physics (HEP) experiments through Transnational Access (TA), we aim to define a data model dedicated to the data management of these user experiments. Our general interest was to research if and how Natural Language Processing (NLP) techniques could be used to provide both support and automatic evaluation of the formal requests for experiments submitted by interested institutions and companies.
This may include the generation of practical pieces of advice (i.e., any sort of ``pre-reviews'') provided to the reviewers or recommendations for the beneficiaries even before their proposal is fully submitted.

Unfortunately, the number of the already evaluated RADNEXT submissions and corresponding reviews is limited for now
, and as such, this data wouldn't be sufficient to ensure the proper training of any machine-learning-based model. Indeed, numerous samples of peer reviews and representative cross-sections of scores are critical to minimize any bias possibly existing in the documents and avoid high-frequency pattern generation. Otherwise, various weaknesses related to artificial text would be visible in the generated reviews, rendering them useless (\citep{yuan2022can}). 

This is to alleviate the scarcity of training data for our project that we envisioned taking advantage of the budding {Open Peer Review} (OPR) movement~ (\citep{ross2017open}). This radical transformation of the scientific assessment of research papers by conference and journal editors, moving from the still prevalent use of blind and secretive reviews to an open approach, is bound to produce large and publicly available resources (papers, reviews, scores, etc.) whose content is clearly related to the experiment proposals we want to help assess within the RADNEXT project. 

Yet, as already mentioned, reliable training data is crucial to help assess whether an ``intelligent'' system capable of providing scientific recommendations and reviews for the RADNEXT community can be designed and implemented. At the same time, since the OPR movement is still in its infancy, it is currently difficult to identify sources that can provide generally accessible reviewing information. Despite the growing number of journals applying an OPR process (20 publishers and 174 journals in 2018 compared to 38 publishers and 617 journals in 2020, according to~\citep{wolfram2019open} and ~\citep{wolfram2020open}), access to these reviews is often restricted, as OPR journals provide authors the option to opt-out of sharing the review process publicly. 

The new Open Review-Based (ORB) dataset introduced here\footnote{\url{https://gitlab.cern.ch/irrad/orb-dataset}} represents a first step into providing both an initial set of publicly available scientific reviews and a computing infrastructure able to accommodate the future introduction of new related sources. In this first version, ORB is populated with the publicly available data provided by the \textit{OpenReview.net} and \textit{SciPost.org} platforms.  If the initial motivation for ORB is the rather specific RADNEXT project\footnote{Incidentally, note that the study of radiation effects is of utmost importance for a society that relies crucially on GPS and telecommunication satellites and other computing-based systems, which can be significantly damaged by a single solar event and its accompanying shower of ionizing particles in the absence of properly designed protection equipment.}, it is quite clear that such a resource will also be of significant value for researchers interested into assessing the impact of the paradigm-shifting OPR movement, both on the academic milieu and the society at large. The Open Peer Review process is at least partially transparent and subjected to a broader community's assessment - therefore improving the quality of the reviews and their objectivity. With a large enough OPR-centered dataset, it will be possible to conduct related research and build NLP models for a possibly more objective document-evaluation process or even checking the presence of possible bias in non-OPR comments. 

In the rest of the paper, we first review the work related to our research in Section~\ref{sec:Related_work}. In Section~\ref{sec:Orb_construction}, we present the ORB dataset design, implementation and main contributions; additionally we motivate their most important features. Potential applications, results of initial illustrative experiments and limitations are presented in Section~\ref{sec:Applications}, before concluding in Section~\ref{conclusion}.  Appendix A provides the ORB datasheet, for reference.

\section{Related work}\label{sec:Related_work}

Having reviewed the Open Review Policy (ORP) notices from MDPI and PeerJ \footnote{Sources:https://mdpi.com/editorial\_process;  https://peerj.com/about/policies-and-procedures}, we found that both publishers support the ORP principles while PeerJ also publishes a peer-review history since February 2023. However, authors have still the possibility to opt-out of making reviews public. 
Even though OPR is encouraged, it is often not trusted enough. Authors and publishers seem to be particularly concerned about their own data privacy. These difficulties result in the relatively small number of available OPR datasets and restricted access to them. For instance, Elsevier offers access to the  "Peer Review Workbench" dataset~(\citep{williams2021}), part of the cloud-based "ICSR Lab" computational platform. Its purpose is to help to advance research evaluation and bring the entire dataset to the end-users via the ICSR platform. However, the official documentation of this dataset\footnote{Source: \url{https://lab.icsr.net/icsr\_lab/workbenches.html}} states that the access is granted only when a proper research proposal is submitted beforehand. This should include title, abstract, expected outcomes, and supporting research.

Some of the most interesting data fields, useful for NLP research (abstract, reviewer's comments), are attached only for accepted articles. We believe that negative and detailed counter-examples are also very valuable input for researchers who would like to perform detailed research on dependencies between scientific papers' content, their reviews, and final decisions.

Based on online research and private discussion with scientific library specialists, we found a few accessible Open Peer Review datasets that supposedly fulfil the criteria that Peer Review Workbench did not.
The first resource is PeerRead~(\citep{kang2018dataset}), a dataset released in 2018. Despite the fact that this dataset was supposed to be released periodically, the last update available in its GitHub repository\footnote{PeerRead, GitHub repository, \url{https://github.com/allenai/PeerRead/}} dates back to May 2018. The content is currently limited, based on a few conferences before 2018.

A second important reference is the Multidisciplinary Open Peer Review Dataset  (MOPRD) by \citep{lin2022moprd}, available online\footnote{MOPRD: A Multidisciplinary Open Peer Review Dataset, \url{http://www.linjialiang.net/publications/moprd/}}.
The authors point out a key fact that we confirmed in our own research: there are very few resources related to open peer-review collections. Additionally, they note that existing  datasets do not cover the whole process of peer reviewing and are mostly based on NeurIPS proceedings\footnote{Conference on Neural Information Processing Systems, \url{https://nips.cc/}}. They compare existing peer-review datasets and their applications; in the vast majority, the existing items are used to predict acceptance and scores~(\citep{gao-etal-2019-rebuttal,kang2018dataset}) or citations~(\citep{plank2019citetracked}).

Expert knowledge is essential in writing reviews, and lack of such domain knowledge in the models generating text is the motivation behind the Meta-Review Dataset (MReD) by ~\citep{shen2022mred}. The authors created a new dataset consisting of 7,089 meta-reviews that was used for structure-controllable text generation. 

The data is retrieved from the  ICLR 2018-2021 conferences. The authors point out that most of the available datasets for summarization were either focused on the news category or contained only a few hundred input papers. The proposed dataset is also annotated by hired professional annotators; as such, the annotation process seems to be done manually.

All the data sources presented often gather abstracts and decisions, or abstracts and comments, or reviews only, while our main goal is to collect the data for multi-purpose experiments with special attention towards automatic assessment and, possibly, to use it in the field of High-Energy-Physics experiments. Therefore, our aim is to focus on datasets that are complete in the sense of providing full-text submissions, reviews' data and metadata.
Moreover, existing datasets are limited to well-established computer-science conferences, with only MOPRD attempting to use multidisciplinary texts from PeerJ Publishing. All three datasets are strongly pre-processed by text segmentation, exclusions of parts of the submissions and data annotations. The annotation is done with external tools or by human experts; it is not an inherent feature derived from the source data. This approach may restrict the experimentation range with the dataset. We believe a better methodology is to provide both the original source data and an interface to perform the pre-processing on demand, when a dataset user wishes to do that. This is the approach advocated here with ORB.

\section{ORB dataset construction} \label{sec:Orb_construction}
In this section we present the design and properties of the ORB dataset. We describe the source of the data, the methodology involved in the structure design and additional resources. ORB is not only a data resource, but also an entire API-like approach to data collection based on recognized data-management processes. 

\subsection{Peer-review data sources} \label{sec:data_source}
\paragraph{OpenReview.net} As first source of data, we chose the \textit{OpenReview.net} website\footnote{\url{https://openreview.net/}}. As it promotes Open Science (including Open Peer Review, Open Publishing and Open Access), we selected it to build the foundations for the ORB dataset. \textit{OpenReview.net} supports an access API, 
which is a REST-like tool to (1) automate the management of user's profiles, submissions, venues and reviews, and (2) provide read access to all the published data.
We used this API to pull the reviews-related data. After content examination, we noticed the following weaknesses.
\begin{enumerate}
    \item There is a lack of a constant data structure or schema, since submissions, reviews, meta-reviews, decisions are represented as one type called "Note".
    \item Every "Note" is represented as a Python dictionary, the keys (densely nested) being set only if values exist, which results in missing keys when parsing data.
    \item Every value exists as a String-type, including "numerical" scores that are concatenated with descriptions.
    \item The process of obtaining data requires a diligent configuration of the client and of its search parameters.
\end{enumerate}
The streamlined design of the ORB dataset is an effort to address and solve these issues.
\paragraph{SciPost.org}  SciPost Foundation, Inc. is a nonprofit and charitable organization that aims to provide 
an online-publication portal. The tool is subscription-free and provides two-way open access to Sci-Post's online journals, papers submissions and reviewing process.
As the "SciPost Terms and Conditions"\footnote{\url{https://scipost.org/terms\_and\_conditions}} state, all online content published on the \textit{SciPost.org} is licensed under license \textit{Creative Commons Attribution 4.0 International (CC BY 4.0)} that allows for redistribution, transforming and building upon the existing material.
SciPost's reports (the name used for a review of a submission) are well structured. They are separated into four sections: (1) Strengths, (2) Weaknesses, (3) Report comments and (4) Requested changes. The total scoring consists of six separate grades describing each submission's Validity, Significance, Originality, Clarity, Formatting and Grammar.
However, the major disadvantage, in comparison to \textit{OpenReview.net}, is the lack of any API allowing data access. All the data is available only as a website; in order to obtain relevant information, the data needs to be extracted from every single HTML page and represented as a data object. 
ORB will facilitate access to SciPost's submissions by including them in the ORB dataset and delivering tools to process the raw HTML and update the dataset in the future.

With the two mentioned sources of data, ORB stores currently over 36,000 submissions (including corrected re-submissions) and 89,000 reviews. Detailed statistics are shown in Table~\ref{statistics-table}.
Additionally, we plan to provide inclusion of new data sources as long as they will follow Open Peer Review principles.

\begin{table}
  \caption{Statistics of data sources and their raw (unprocessed) peer-review deliverables. The number of submission represents unique submissions (re-submissions are not counted in this case).}
  \label{statistics-table}
  \centering
  \begin{tabular}{lrrrr}
    \toprule
    Source     & Unique submissions  & \makecell{Submissions with\\ multiple revisions}& \makecell{Submissions\\with reviews}  & \makecell{Number\\of reviews} \\
    \midrule
    OpenReview.net & 34,030 & - - - &24,709 & 85,470     \\
    SciPost.org     &  2,919 &2,081 &2,801& 7,409      \\
    \midrule
    Total & 36,949 & 2,081& 27,510& 92,879\\
  \end{tabular}
\end{table}

\subsection{Dataset structure design}
\begin{figure}[h]
\includegraphics[width=\textwidth]{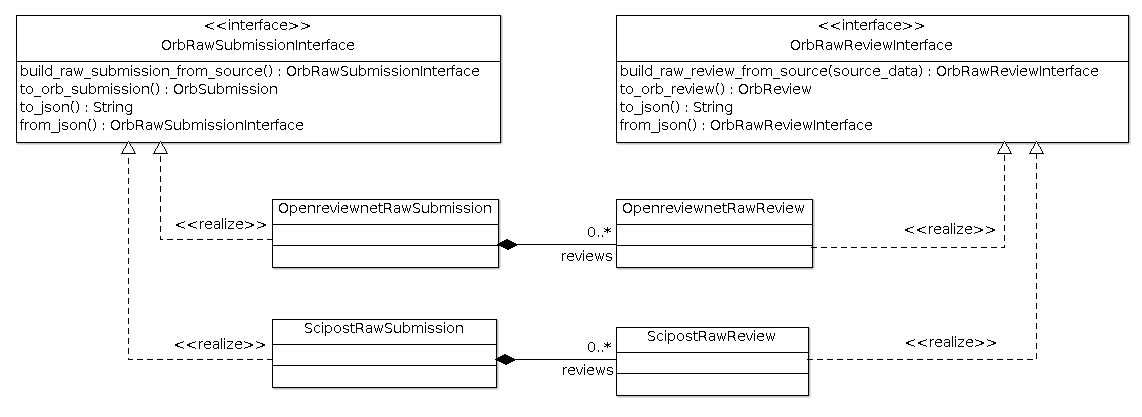}

\caption{UML diagram presenting the predefined OrbRaw* interfaces and their implementation for existing data sources (\textit{OpenReview.net} and \textit{SciPost.org}). This general view (without class fields and methods) is provided to illustrate the architecture of the abstraction layer. Its implementations are wrapping up the raw data either delivered from an API as lists and dictionaries or from HTML source. They are organized as containers; the submission and review classes, whose properties are simple strings, are intended to be converted later into more meaningful representations.}

\label{orb-interfaces-uml}
\end{figure}

Due to the observed weaknesses of the existing datasets and disadvantages of using strongly pre-processed data sources, we designed a structured approach to deliver the data and provide a way for ORB-dataset users to extend it if they wish to do so.
Additionally, we intend to implement proper methods that would clean or process the data - but according to the desired use and only when necessary.
\paragraph{Predefined Interfaces} Currently, we use two separate data sources. However, the aim of ORB is to deliver means to extend the dataset in the future.
Different sources of data will differ not only in content but also in their structure or have no defined structure at all - similarly to  about 90\% of the world's data that is particularly challenging to integrate~(\citep{mehmood2019implementing}). With the introduction of two interfaces representing (1) the submission (\texttt{OrbRawSubmissionInterface}) and (2) the associated review (\texttt{OrbRawReviewInterface}), we aim to provide the minimum needed to be implemented for present and future usage of these data sources. We refer to them as "\textbf{OrbRaw*}" interfaces (see Figure~\ref{orb-interfaces-uml}).

\paragraph{Raw data representation} We observed that the submissions and reviews from different online libraries, OPR journals and submission portals host various data and metadata. Their designs usually follow similar ideas, but details often do not correspond to one another when comparing two sources. To properly accommodate them and provide a seamless transition to ORB dataclasses, each one should implement the corresponding OrbRaw* interfaces (see Figure~\ref{orb-interfaces-uml} that presents the ORB classes structure and relationships in UML format). 
As the part of the ORB infrastructure, we provide implementations of OrbRaw* interfaces for the \textit{OpenReview.net} and \textit{SciPost.org} data collections. The fields of these classes reflect the original information stored on the data portals. With their own implementation of the defined abstract methods, they fit the concept of ORB by design.

\paragraph{ORB dataclasses} The most important part of our work are target ORB classes that represent submissions and reviews in a more detailed and structured way. We defined them as "dataclasses" responsible exclusively to store data objects.
Unification of heterogeneous data is a complicated task. However, the analysis of the conceptual schema of each source  results in better data integration and enhanced reusability~\citep{castano2001global}. Such data integration provides, in return, better accessibility, especially in case of data from non-cooperating sources~\citep{stupnikov2021applying}.

In comparison to OrbRaw*, the \textbf{Orb*} classes are aimed to represent collections that simplify browsing and perform unification of the originally heterogeneous data, with a possibility to introduce new data sources in the future.
To achieve such a goal, we introduced the new \textbf{CEDIgaR} procedure that can be followed in order to perform continuous data integration in any extensible dataset.
\begin{enumerate}
    \item \textbf{Create} a schema to provide conceptual representation using domain knowledge (UML diagram or ontology), to form a baseline for dataset design.
    \item \textbf{Establish} and select a (new) source of data, e.g., existing datasets, HTML web pages, databases.\label{new-datasource}
    \item \textbf{Define} a transitional structure corresponding to the new data source.
    \item \textbf{Investigate} the differences between the existing schema (baseline) and the incoming data, and then: 
        \begin{enumerate}[(a)]
            \item \textbf{generalize} the new data source into an existing baseline and
            \item \textbf{accommodate} the important changes and extend the baseline structure if any important information was lost in previous sub-step.
        \end{enumerate}
    \item \textbf{Release} a new version of the dataset or/and go to step \ref{new-datasource}) and repeat the procedure to add new data sources.
\end{enumerate}

\begin{figure}[h]
\includegraphics[width=\textwidth]{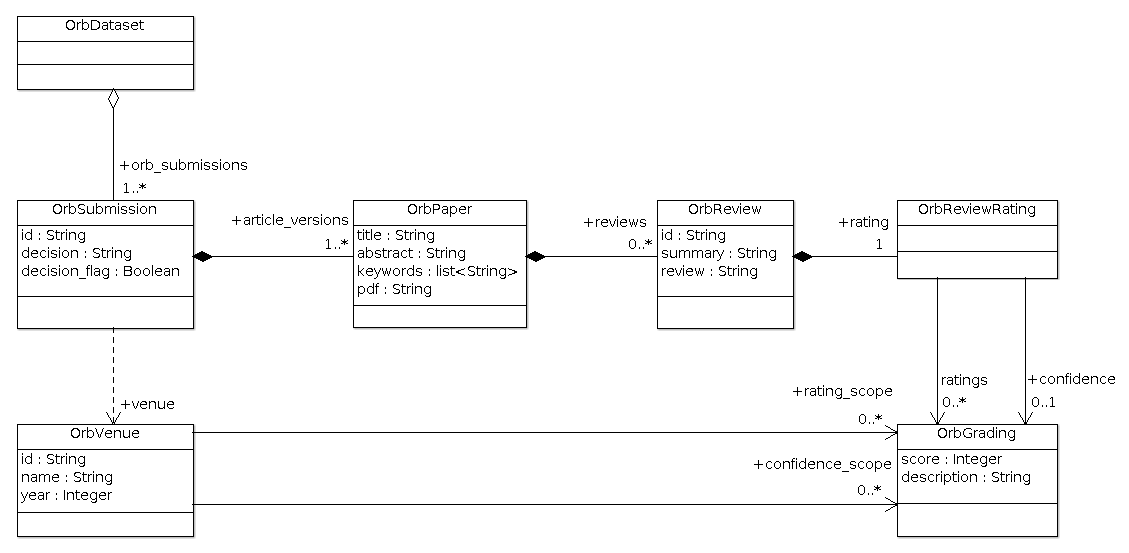}
\caption{Target Orb* dataclasses of the ORB framework. 
All the OrbSubmissions of OrbPapers for a given OrbVenue (conference, journal, etc.) are collected as an OrbDataset. Each paper, associated to its \texttt{title},  \texttt{abstract} and URL of its \texttt{pdf} content, is associated to zero or more OrbReviews. Each includes a textual \texttt{review} for the paper and one OrbReviewRating, defined, in terms of scope and confidence, according to the OrbGrading scheme specific to the OrbVenue. The \texttt{decision} taken by the venue committee for each paper is stored in each OrbSubmission.
This foundational UML diagram is open to enhancement in terms of generalization or further extensions when additional data sources will be handled.} \label{orb-dataset-classes-uml}
\end{figure}

Following the presented steps, we created the target structure of ORB, a dataset for Open Peer Review processes that currently includes the submissions from two data sources (see Figure \ref{orb-dataset-classes-uml}).

\subsection{Data architecture}
In this paragraph, we provide a description of the ORB data architecture, which we consider as a process defining a complete workflow, starting from data collection, through its processing and necessary transformations, up to the point of its final delivery to be used~\citep{IBM2022architecture}.
To maintain clarity and extensibility of our solution, the data architecture is implemented in the framework of ETL (Extract, Transform, Load) processes, a concept designed for data extraction, processing and exchange~\citep{vassiliadis2009survey} and widely used in data modeling and warehousing~\citep{trujillo2003uml}.The ETL process is a three-step method to facilitate data collection, often from more than one single source. These consecutive steps \textbf{extract} the data (e.g., using a dedicated API or HTML scraper), then \textbf{transform} the data into a format that can be queried, and finally \text{load} the target structures in the computer memory or data warehouse~\citep{el2011proposed}.

Data pre-processing, unification and generalization can be built directly into modern data pipelines and machine learning dataset interfaces~\citep{lubbering2022datastack}, which help to mitigate redundant code and recurring implementation steps.

\begin{figure}
\centering
\includegraphics[width=\textwidth]{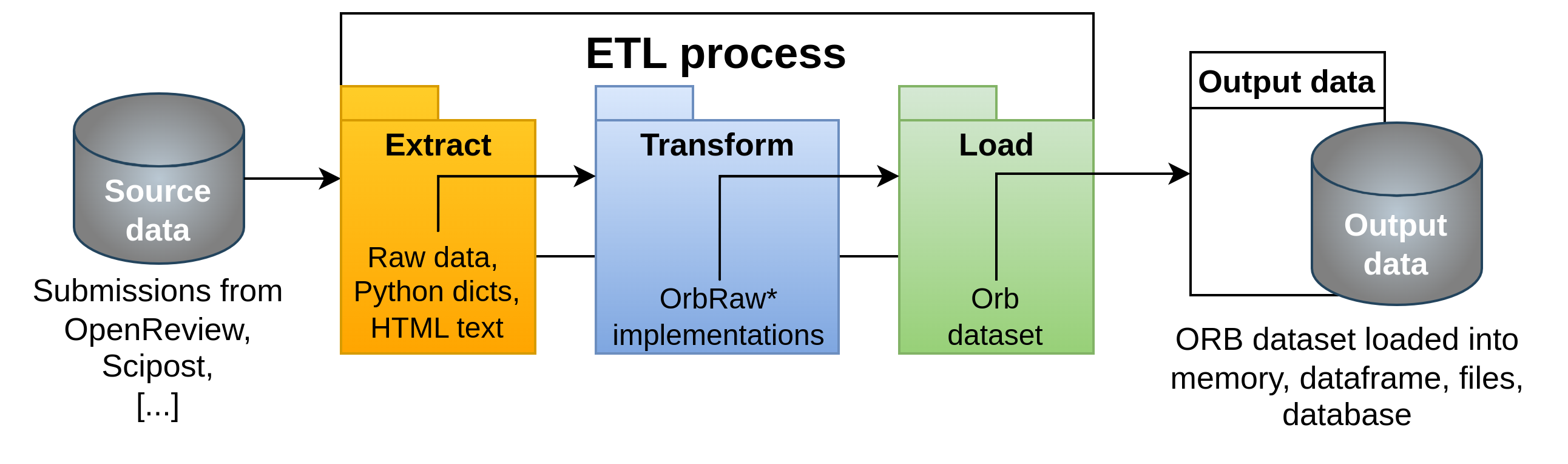}

\caption{ETL process and data representations during its consecutive steps. Data is extracted from datasources in its original form, then transformed into OrbRaw* objects (that aim to keep their original concept but in a structured way) and finally loaded as Orb* dataclasses.}

\label{etl-drawio}
\end{figure}

We built our own ETL processes and integrated them within the developed ORB classes (see Figure \ref{etl-drawio}). That allows for providing a reusable approach to dataset updates; re-running the ETL will download information that is up-to-date and refresh the submissions and reviews. 

\subsection{Delivered resources} \label{sec:delivered_resources}
In summary, the ORB project led to the definition and development of the following original research contributions, which are publicly available for usage and further development.
\begin{itemize}
    \item The data files containing the instances of all the current collected submissions and reviews, structured according to the ORB principles.
    \item The OrbRaw* interfaces (see Figure~\ref{orb-interfaces-uml}) and Orb* dataclasses (see Figure~\ref{orb-dataset-classes-uml}) that represent the Open Peer Review process.
    \item A set of reusable ETL processes and programs dedicated to the management of dataset updates.
    \item The OWL-based \texttt{orb.rdf} ontology for ORB classes and relations, which, in addition to its current formal specification purpose, is intended to be connected to the ORB dataset to develop a Knowledge Graph for Open Peer Review data.
\end{itemize}
In addition to these concrete artefacts, we introduced the concept of {CEDIgaR}, a methodology specifying the key steps required for designing a dataset subject to continuous development.

\section{Applications and limitations} \label{sec:Applications}
In this section, we briefly discuss the potential applications that Open Peer Review data, in general, and the ORB dataset, in particular, enable. Initial illustrative experiments with ORB using NLP techniques are described and the results are presented. Finally, we discuss the limitations that such data present, both on technical and ethical grounds.

\subsection{Overview of Open Peer Review data applications}

Using ORB, we aim to represent the key data elements of the peer-reviewing process. This kind of data covers a variety of scientific and technical applications concerning analysis with NLP techniques, text generation, but also more social aspects of this process such as ethics in scientific activities. Nevertheless, we focus here on the more technical side of potential applications.

The Open Peer Review process provides interesting possibility to link submissions and reviews with indirect metadata~\citep{wang2023have}, e.g., the information that can be associated with authors, or with the paper itself that was published as a preprint even before submission to journal/conference supporting Open Peer Review.
As the ORB dataset contains a structured representation of (re)submissions and their history concerning reviews, requested changes and corrected papers, we believe it provides valuable data that can be used for text summarization, automatic review generation or even text annotations.
Of course, in certain cases, the ORB records will have to be pre-processed accordingly.

An original strength of ORB is the inclusion of papers and reports in various fields of Physics (including those using experimental approaches). That allows to train machine learning models to learn semantic relationships between words in texts from disciplines different from those commonly used in Computer Science. Additionally, it can improve such specific tasks as learning embeddings and knowledge extraction in the discipline of Physics~\citep{shang2022representation}, domain in which the proper resources describing research-related knowledge seem to be very uncommon.

\subsection{Preliminary illustrative experiments}
\label{sec:preliminary}
Based on this first version of ORB, we were able to perform some preliminary experiments that illustrate the kind of analyses that can be designed thanks to this new dataset. More specifically, we aim to predict paper acceptance based on its textual embeddings and the grades with given confidence. These results are briefly presented here.

To perform these initial experiments, we trained the Paragraph Vector-based model~\citep{le2014distributed} (known also as Doc2Vec) using the \textit{scientific\_papers} dataset~\citep{cohan2018discourse}, which contains abstracts and texts of 200,000 instances of ArXiv and PubMed open-access documents, to learn proper scientific-related corpus embeddings.
Training was performed online on the SWAN (Service for Web based ANalysis)\footnote{The Swan Service, \url{https://swan.cern.ch}} portal (4 CPU cores, 16 GB of RAM) with usage of Jupyter Notebooks.
The Doc2Vec model was configured as a distributed memory model of Paragraph Vector with hierarchical softmax for model training. The size of feature vectors was set to 512 elements.

Using the PDF documents' URLs from the ORB dataset, the full-text content of each paper was extracted using the Python package \textit{pdfplumber\footnote{pdfplumber's GitHub repository, \url{https://github.com/jsvine/pdfplumber}}}.  Based on the embeddings inferred from the above described model, we could predict the acceptance of the submitted papers present in the ORB dataset. In this experiment, we used data from the SciPost subset only, as a similar experiment for OpenReview was already conducted in works referenced in Section~\ref{sec:Related_work}.
The total set of 4,917 SciPost elements (2,691 original submissions with corrected re-submissions) was split in two train and test subsets, with a 0.8:0.2 ratio.

\paragraph{Acceptance prediction} Using a Support Vector Classifier with a Radial Basis Function (RBF) kernel and the k-Nearest Neighbors algorithm (\texttt{n\_neighbors} = 10), we obtain the confusion matrices for paper acceptance presented in Tables~\ref{confusion-matrix} and \ref{confusion-matrix-knn}, where $p$ stands for "accepted" and $n$ for "rejected" (with a prime when dealing with predictions). In this simplified proof-of-concept experiment, the accuracy score obtained by SVC is 0.59, while the one for kNN is 0.55. This somewhat poor result may be due to one inherent feature of SciPost data: versioning. There, the corrected text between successive versions may not change too much, so, for very similar embeddings, we are enforcing the model to learn two different decisions, which negatively affects the learning process. 

\begin{table}[h]
\parbox{.40\linewidth}{
\caption{Confusion matrix for SVC.}
  \label{confusion-matrix}
\centering
\noindent
\renewcommand\arraystretch{1.5}
\setlength\tabcolsep{0pt}
\begin{tabular}{c >{\bfseries}r @{\hspace{0.7em}}c @{\hspace{0.4em}}c @{\hspace{0.7em}}l}
  \multirow{10}{*}{\rotatebox{90}{\parbox{1.1cm}{\bfseries\centering PREDICTION}}} & 
    & \multicolumn{2}{c}{\bfseries ACTUAL} & \\
  & & \bfseries p & \bfseries n & \bfseries total \\
  & p$'$ & \MyBox{577}{} & \MyBox{6}{} & 583 \\[2.4em]
  & n$'$ & \MyBox{400}{} & \MyBox{1}{} & 401 \\
  & total & 977 & 7 &
\end{tabular}
}
\hfill
    \parbox{.45\linewidth}{
\caption{Confusion matrix for kNN.}
  \label{confusion-matrix-knn}
\centering
\noindent
\renewcommand\arraystretch{1.5}
\setlength\tabcolsep{0pt}
\begin{tabular}{c >{\bfseries}r @{\hspace{0.7em}}c @{\hspace{0.4em}}c @{\hspace{0.7em}}l}
  \multirow{10}{*}{\rotatebox{90}{\parbox{1.1cm}{\bfseries\centering PREDICTION}}} & 
    & \multicolumn{2}{c}{\bfseries ACTUAL} & \\
  & & \bfseries p & \bfseries n & \bfseries total \\
  & p$'$ & \MyBox{498}{} & \MyBox{85}{} & 583 \\[2.4em]
  & n$'$ & \MyBox{361}{} & \MyBox{40}{} & 401 \\
  & total & 859 & 125 &
\end{tabular}
}
\end{table}

\paragraph{Grading statistics prediction} The second experiment aimed to predict the statistics related to the submissions final scores, i.e., the averages of the final and confidence scores with their respective variances for a paper. This approach has been chosen to investigate the impact of the varying number of reviews for submissions.
To achieve this goal, we selected another subset of the ORB dataset: the venues that had a similar grading scope of the final score (range similar to 1--10) and confidence (range 1--5). This choice allowed us to use the largest number of submissions.
We ended up with over 16,000 submissions from conferences such as NeurIPS and ICLR but also UAI (Uncertainty in Artificial Intelligence).\\
We used the \texttt{keras\_tuner} package\footnote{Keras documentation: KerasTuner, \url{https://keras.io/keras_tuner/}} to define the trainable hyper-model of the prediction neural network. With automatic tests of various networks' parameters (such as optimizer choice, number of hidden layers, etc.), the tuning result yields the model with the  configuration in Table~\ref{tbl:hyper}.
\begin{table}[h]
  \caption{Result of hyperparameter tuning for neural network predicting grading statistics.}
  \label{hyperparameter-table}
  \centering
  \begin{tabular}{lc}
    \toprule
    Hyperparameter     & Value \\
    \midrule
    Optimizer           & Adam \\
    Hidden Layers       & 8 \\
    Neurons per layer   & 150 \\
    Activation function & ELU \\
  \end{tabular}
  \label{tbl:hyper}
\end{table}
We performed an MAE (Mean Absolute Error) metric evaluation to get an idea of how large the differences between the actual data and the model predictions are (see Table~\ref{grading-statistics-table}). These values show that the network prediction may be off by around 0.8 per final score and 0.4 per confidence.
\begin{table}[h]
  \caption{Values of MAE (Mean Absolute Error) for the final and confidence scores and their variances}
  \label{grading-statistics-table}
  \centering
  \begin{tabular}{rrrr}
    \toprule
    Score error &  Score variance error  & Confidence error & Conf. variance error \\
    \midrule
    0.87 & 0.78 & 0.40 & 0.30    \\
  \end{tabular}
\end{table}
One possible reason for these somewhat wide error ranges is that this experiment does not take into account that the OpenReview grades are often influenced by  elements other than the mere submitted text. For instance, an important factor is the impact of the possible discussions between authors and reviewers. 

To conclude on these preliminary experiments, it is clear that improving their results would require a more thorough analysis of the underlying data, which we leave as future work, our goal here being only to give a feeling for the type of uses we think ORB could be suitable for. In any case, these preliminary experiments already show a very important feature coming with the ORB dataset: the possibility to create quite large and task-specific subsets of review data for more focused analysis, as shown in the grading-prediction experiment.

\subsection{Limitations}

The preliminary experiments described in Section~\ref{sec:preliminary} provide confidence in the operational usability of the ORB framework while illustrating very simple use cases for the ORB dataset, thus paving the way to more advanced, NLP-based analyses of this corpus of text, which could have a significant impact for the Open Science research. 

Yet, we noticed the following limitations and issues with these data sources. Firstly, the papers' full-texts are delivered as PDF files. The PDFs are not stored in ORB, just their URLs, in order to keep the dataset within reasonable size limit. We do not address the issue of PDF data extraction in the framework of this paper, as it poses multiple questions regarding the segmentation of such data and the possibility to design it as multi-modal process (e.g., text, figures, tables, metadata). 

Secondly, no personal data were stored in the original data sources during the peer-review process (neither for the authors nor for the reviewers). However, the ORB dataset, as also the used data sources, contain the papers' titles and URLs to PDF. This naming information could thus be indirectly inferred or via PDF extraction and analysis, possibly posing ethical privacy concerns. 

Ethics is, indeed here, an important possible concern for the type of data collected in ORB, as the recent development of the field of Digital Ethics shows \citep{veliz2021}. The OpenReview and Scipost websites already deal with some of the these concerns: the submissions and their reviews are envisioned to be public, a fact made quite clear to the contributors by the sites' guidelines. Yet, the type of large-scale cross analysis that NLP techniques and large-language models enable might require this consent to be more informed. For instance, already now, in the case of a rejected paper, agreeing to still take part in the open-access process is up to the authors (an additional opt-in decision).

Finally, and still in the realm of ethics, the provided data (especially reviews) are in an unchanged form, being straight copies from those present in the original provider sources. Therefore, no semantic examination of the about 89,000 reviews for the presence of offensive content was performed. Relying on the scientists' code of ethics involved during the review process might not be enough in the long term, especially in the presence of possible external hacks. Dealing with such issues at a large scale and with the specifics of the scientific domain could be an interesting research endeavor.

\section{Conclusions and future work}
\label{conclusion}
In this paper, we have presented the design and implementation of the ORB dataset, which contains Open Peer Review process data built from heterogeneous data sources, its infrastructure and some illustrative analysis use cases. It contains scientific submissions and their reviews from established conferences in the field of Computer Science (\textit{OpenReview.net}) and from open journals in the field of natural sciences, especially High-Energy Physics (\textit{SciPost.org}). To the best of our knowledge, ORB is the first dataset containing experimental HEP-related peer review data.

The concrete results of our work (datafiles, dataset design, interfaces and dataclasses implementation incorporated into ETL processes, ontology) are shared publicly for inspection, usage and discussions regarding further development on the CERN GitLab server at \url{https://gitlab.cern.ch/irrad/orb-dataset}.
Moreover, to design the ORB dataset for continuous development, as needed by the current fluid situation of the Open Science movement, we introduced and followed the new {CEDIgaR} methodology, another contribution of this work.

We foresee periodical updates of the ORB dataset and infrastructure and plan on extending it with new Open Peer Review data sources. Each extension will come with necessary changes in the design (according to CEDIgaR) and corresponding implementations of dataclasses.
Additionally, we plan to use the ORB dataset in the framework of the RADNEXT project to support automatic assessment of research proposals. It should be beneficial for both applicants and reviewers as these activities may involve the pre-review generation of comments or recommendations, even before experimental proposals are submitted.

\begin{ack}
This project has received funding from the European Union's Horizon 2020 research and innovation programme under grant agreement No. 101008126.
For advice regarding the OpenScience approach and for recommending SciPost, we thank our colleagues working at CERN:  Micha Moskovic, manager of \textit{Inspirehep.net}\footnote{INSPIRE-HEP is an open-access digital library for high-energy physics (HEP) \url{https://inspirehep.net/}.}, and  Kamaran Naim, Head of Open Science.
\end{ack}

{
\small
\bibliographystyle{plainnat}
\bibliography{references}



\appendix
\renewcommand\thefigure{\thesection.\arabic{figure}}  
\setcounter{figure}{0}    
\renewcommand\thetable{\thesection.\arabic{figure}}  
\setcounter{table}{0}    

\newpage
\section{Datasheet for the ORB dataset}
\subsection{Motivation}
\label{sec:Motivation}
The main motivation behind creation of ORB dataset (Open Review-Based) is usage for automatic comments and reviews generation for proposals in the field of High-Energy Physics (HEP).
Since (to our best knowledge) the existing dataset are mostly limited to the discipline of Computer Science, we decided to build our own resource that contains data of Open Review process in HEP (see SciPost datasource).
\subsection{Composition}
\label{sec:composition}
The dataset consists of the data files (binary files and JSON-based format) containing the instances of all  collected submissions and reviews, structured according to the ORB principles. Currently it uses the OpenReview.Net and SciPost.org as sources of Open Peer Review data.
Besides included datafiles, the dataset delivers the Python implementation of OrbRaw* interfaces and Orb* dataclasses that represent the Open Peer Review process, with history and changes when available.

Orb* dataclasses are highly structured generalization that is supposed to unify different data sources (see Figure~\ref{orb-dataset-dataclasses}).

\begin{figure}[h]
\centering
\includegraphics[width=\textwidth]{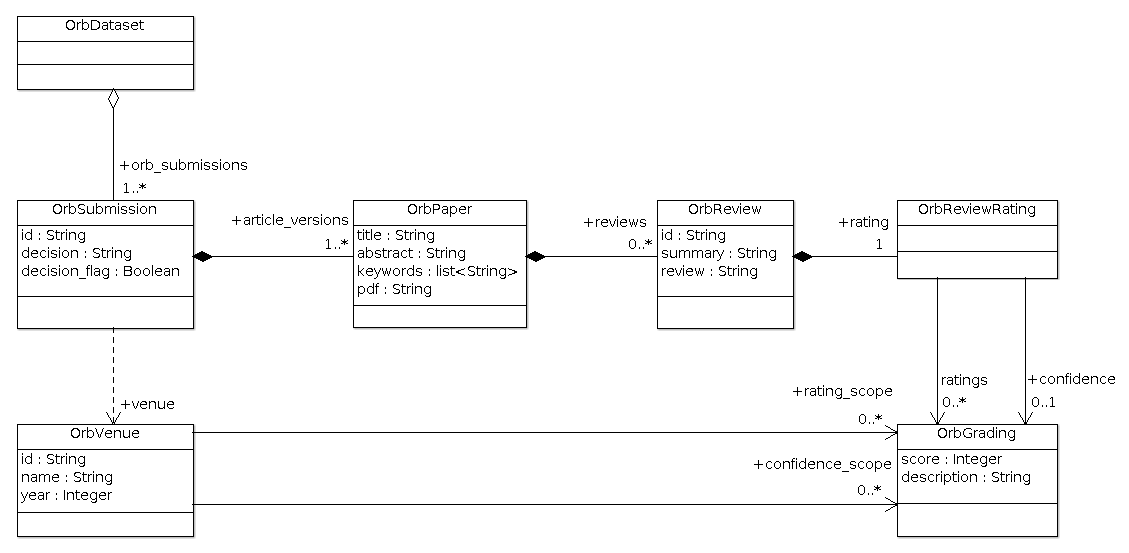}
\caption{ORB dataset UML diagram presents the layout of target dataset and relationship between classes.} \label{orb-dataset-dataclasses}
\end{figure}

\subsection{Collection process}
The dataset instances are collected from two sources, however, the methods for data collection differs in these cases:
\begin{enumerate}
    \item OpenReview.net: OpenReview delivers implementation of API client that we used to gather the papers submission and their review in form of "notes",
    \item SciPost.org: it is a web portal that contains submissions and reviews in HTML format, therefore we implemented own scrapper to extract relevant data from individual webpages.
\end{enumerate}
For both cases, we intend to publish the examples of API usage and implementation of our own SciPost scrapper.
The collection process is implemented as the ETL process (Extract - Transform - Load), allowing for potential users to re-run it and update the dataset with Python script.
All stages of data collection (collected raw information, OrbRaw* instances and target Orb* instances) can be stored for further use.  

\subsection{Preprocessing}
We avoided the unnecessary preprocessing that strips information in order to deliver all relevant data properties. However, the collection process is aiming to unify data, therefore ORB* classes are reorganizing the data format in order to fit the schema described in the Section~\ref{sec:composition}.
With addition of new data sources we expect further schema modification, beyond the status that is currently presented.
\subsection{Potential uses}
ORB dataset contains representation of the Peer Review process, therefore its greatest value is in the papers full-text and corresponding reviews. However, we designed it to deliver supplementary data, e.g., paper submission history, data properties such as paper acceptance status or numerical scores representation.

We envision the possibilities of following applications such as:
\begin{itemize}
    \item Large Language Models (LLMs) training and fine-tuning,
    \item embeddings training,
    \item reviews and comments generation related to Peer Review Process and scientific assessment,
    \item scientific texts summarization,
    \item automatic classification and quality assessment of scientific papers,
    \item sentiment analysis of Peer Review comments,
    \item statistics generation related to the paper-submission process,
\end{itemize}

\subsection{Distribution}
We are building the ORB dataset upon available services and data: OpenReview (open access, MIT License\footnote{The MIT License, \url{https://opensource.org/license/mit/}} specified for API client) and SciPost (open access, Creative Commons Attribution 4.0 International (CC BY 4.0) License\footnote{Attribution 4.0 International (CC BY 4.0), \url{https://creativecommons.org/licenses/by/4.0/}}).

To facilitate usage of ORB, its relevant code and other deliverables, we chose licensing under the \textit{Creative Commons Attribution 4.0 International (CC BY 4.0) License}. The copyright remains authors property, however, we agree to sharing and adaptation according to specified license terms.

\subsection{Maintenance}
We are planning periodical dataset update and maintenance in order to accomplish the goals specified in Motivation section (refer to Section ~\ref{sec:Motivation}). More precisely, we plan
\begin{itemize}
    \item updates, i.e., reruning the ETL processes to download, process and store the new submissions that arrived since the previous datafiles publication, and 
    \item maintenance, with possible bugfixing and addition of new features, including th definition of new datasources and relevant code implementation.
\end{itemize}

Additionally (as specified in the previous section), the code is publically available on CERN Gitlab and licensed under Creative Commons Attribution 4.0 International (CC BY 4.0) License. Therefore everyone willing will have access to accommodate the dataset to their use and perform updates.

We plan setting up an additional repository to coordinate git pull requests. More information will be available in the Readme file on CERN gitlab (see Section
~\ref{sec:introduction}).
}

\end{document}